\documentclass[
fleqn,usenatbib]{mnras}

\usepackage{newtxtext,newtxmath}
\usepackage[T1]{fontenc}
\usepackage{colortbl}
\usepackage{xcolor}
\DeclareRobustCommand{\VAN}[3]{#2}
\let\VANthebibliography\thebibliography
\def\thebibliography{\DeclareRobustCommand{\VAN}[3]{##3}\VANthebibliography}

\usepackage{graphicx}	% Including figure files
\usepackage{amsmath}	% Advanced maths commands
\usepackage{xcolor}
\usepackage{subcaption}
\graphicspath{{picture/}}
\usepackage{pgffor}
\usepackage{ulem}

\title[Point-source simulation calibrated by LoTSS]{A New Model for the Extragalactic Radio Sky at Low Frequency Calibrated Using the LOFAR Two-metre Survey}

\author[Jinyang Lin et al.]{
Jinyang Lin,$^{1}$
Zhenghao Zhu,$^{2}$\thanks{zhenghao@shao.ac.cn}
Renyi Ma,$^{1}$\thanks{ryma@xmu.edu.cn}
Anna Bonaldi,$^{3}$
Huanyuan Shan$^{2, 4, 5}$\thanks{hyshan@shao.ac.cn}
\\
$^{1}$Department of Astronomy and Institute of Theoretical Physics and Astrophysics, Xiamen University, Xiamen, Fujian 361005, China\\
$^{2}$Shanghai Astronomical Observatory, Chinese Academy of Sciences, 80 Nandan Road, Shanghai 200030, China \\
$^{3}$SKA Observatory, Jodrell Bank, Lower Whitington, Macclesfield, SK11 9DL, UK\\
$^{4}$Key Laboratory of Radio Astronomy and Technology, Chinese Academy of Sciences, A20 Datun Road, Chaoyang District, Beijing, 100101, P. R. China\\
$^{5}$University of Chinese Academy of Sciences, Beijing 100049, China
}

\date{Accepted XXX. Received YYY; in original form ZZZ}

\pubyear{2015}

\begin{document}
\label{firstpage}
\pagerange{\pageref{firstpage}--\pageref{lastpage}}
\maketitle

\begin{abstract}
Building the radio sky template are crucial for detecting the 21 cm emission line signal from the Epoch of Reionization (EoR), as well as for other cosmological research endeavors. Utilizing data from the LOFAR Two-meter Sky Survey (LoTSS) at 150 MHz, we recalibrated the luminosity function for various types of radio sources, including High Excitation Radio Galaxies (HERGs), Low Excitation Radio Galaxies (LERGs), Radio-Quiet Active Galactic Nuclei (RQ-AGNs), and Star-Forming Galaxies (SFGs). We subsequently updated the Tiered Radio Extragalactic Continuum Simulation (T-RECS) code to generate refined mock radio source catalogues.
The simulated source counts from this work align more closely with observed data at redshifts greater than $z>4$. Additionally, the differential source counts in total intensity within the flux density range of $0.1-1~\mathrm{mJy}$ closely mirror actual observations. 
Due to our model incorporating a lower number of faint sources compared to T-RECS, it predicts a reduced power spectrum for point sources, suggesting a potential advantage in studies in low frequency band.

\end{abstract}


%%%%%%%%%%%%%%%%%%%%%%%%%%%%%%%%%%%%%%%%%%%%%%%%%%

%%%%%%%%%%%%%%%%% BODY OF PAPER %%%%%%%%%%%%%%%%%%

\section{Introduction}

As well known, intensity maps of 21\,cm emission line at redshift around z $\sim$ 6 and beyond provide us with an important and unique probe to investigate the physics at the Epoch of Reionization  \citep[EoR,][]{2012RPPh...75h6901P}.  However, the reionization signal encounters significant challenges from systematic contamination due to foreground radiation, i.e., the local emission from the Milky Way, the diffuse and compact extragalactic sources \citep{2002ApJ...564..576D, 2008MNRAS.388..247D, 2009ApJ...695..183B, 2008MNRAS.389.1319J, 2010MNRAS.409.1647J, 2018MNRAS.479..275S, chapman2019foregrounds}.

%However, the reionization signal encounters significant challenges from systematic contamination due to foreground radiation \citep{2002ApJ...564..576D, 2008MNRAS.388..247D, 2009ApJ...695..183B, 2008MNRAS.389.1319J, 2010MNRAS.409.1647J, 2018MNRAS.479..275S, chapman2019foregrounds}, instrumental effects, and other distortions such as the effects of ionosphere \citep{2017MNRAS.471.3974J}. Foreground radiation includes various non-cosmological radio emission sources within the relevant low-frequency range that can affect the measurements. This encompasses diffuse synchrotron radiation and free-free emission originating from \textbf{the} Milky Way, as well as extragalactic point sources. The brightness of the foreground radiation could exceed that of the crucial EoR signal by four to five orders of magnitude \citep{2010ARA&A..48..127M}.

The compact extragalactic point sources have been estimated to contribute approximately $27\%$ of the total foreground radiation at $150~\rm MHz$ \citep{1999A&A...345..380S}. Particularly in the context of small-scale regions within the power spectrum, extragalactic point sources emerge as the most important foreground element \citep{2017PASA...34...33P, 2017ApJ...845....7M, 2018MNRAS.479.2767Y}. 

The point sources can be classified as Active Galactic Nuclei (AGNs) and Star Formation Galaxies (SFGs). For non-AGN sources, both the synchrotron radiation emitted by relativistic electrons associated with supernovae, as well as the free-free emission from HII regions, contribute to the radio emission\citep{1985ApJ...298L...7H,1992AIPC..254..629C,2010MNRAS.409...92J}, 
while for AGNs, because of the release of huge gravitational energy as the gas falls towards the central black holes, all the components of the accretion system are capable of producing radio emission. Based on the radio loudness, i.e., the ratio of the radio to optical luminosity, AGNs are divided into radio-loud (RL) and radio-quiet (RQ) types, which constitute approximately 10\% and 90\% of the AGN population, respectively \citep{1989AJ.....98.1195K,Sikora_2007}. 

The radio luminosity of RL AGNs is dominated by the relativistic jets, in which high-energy electrons are accelerated and then emit synchrotron photons in the strong collimating magnetic fields\citep[e.g.][]{2019ARA&A..57..467B}. Considering the geometry of the radio image, it can be further divided into FRI and FRII sub-populations \citep{1974MNRAS.167P..31F}. Considering the different optical and ultraviolet spectral features, which should be due to the different accretion modes that are efficient or non-efficient in radiation, RL AGNs are also divided into HERGs and LERGs \citep{2014ARA&A..52..589H,2020NewAR..8801539H}.
In contrast, the origins of radio emission in RQ-AGNs remain unclear\citep{2019NatAs...3..387P}, though different models have been postulated, including star formation activities like SFGs\citep{1991MNRAS.251P..14S}, the diminutive jets \citep{2001ApJS..133...77H}, the corona\citep{2008MNRAS.390..847L}, and the disk winds\citep{2008ApJ...682L..81W}.

% The observed radio emission from these galaxies often exhibits a strong correlation with their infrared luminosity, which is used to measure the star formation rate, as referenced in the literature \citep{1992AIPC..254..629C,1985ApJ...298L...7H,2010MNRAS.409...92J}, leading to their classification as Star-Forming Galaxies \citep{chapman2019foregrounds}.

%Recent studies on sub-$\mu$Jy radio sources suggest that their composition is influenced by two distinct accretion modes \citep{2006ApJ...642...96E, 2007MNRAS.376.1849H}. The most powerful RL AGNs are primarily driven by highly efficient accretion of cold gas onto geometrically thin, optically thick accretion disks \citep{1973blho.conf..343N, 1973A&A....24..337S}. This mode of accretion gives rise to highly excited emission lines in the host galaxy's spectrum. Consequently, radio sources fueled by this mode of accretion are referred to as HERGs, which typically exhibit accretion rates ranging from $1\%$ to $10\%$ of their Eddington rate \citep{2012MNRAS.421.1569B}. 

%On the other hand, relatively less luminous AGNs belong to the category of LERGs. They are primarily fueled by the inefficient accretion of warm gas from the intergalactic medium onto geometrically thick accretion disks. Their accretion rates typically remain below $1\%$ of their Eddington rate \citep{2012MNRAS.421.1569B}. LERGs typically do not exhibit X-ray emissions associated with the accretion process \citep{2006MNRAS.370.1893H} or infrared radiation originating from obscured nuclear  torus \citep{2006ApJ...647..161O, 2013MNRAS.436.1084M}.

Radio sky simulation is a useful tool in quite a few aspects of astronomy. For example, it can be used to assess the completeness of radio surveys or to predict the population of radio sources that will be observed in given or future surveys; It was applied to test clustering theories of cosmology by generating random samples. Very important to us, in the research of reionization, it provides a way to predict the foreground noise of the 21\,cm signal \citep[e.g.][]{2008MNRAS.388.1335W, 2010MNRAS.405..447W, 2019MNRAS.482....2B, 2023MNRAS.524..993B,2023MNRAS.523.1729B}.

%The simulations of extragalactic radio sources provide us a tool to investigate the extragalactic sky
%\citep[e.g.][]{2008MNRAS.388.1335W, 2010MNRAS.405..447W, 2019MNRAS.482....2B, 2023MNRAS.524..993B}. 
The Tiered Radio Extragalactic Continuum Simulation (T-RECS) \citep{2019MNRAS.482....2B, 2023MNRAS.524..993B} is designed to simulate two primary populations of radio galaxies, i.e., AGNs and SFGs. It takes into account not only the luminosity functions for different redshift, but also the clustering properties \citep{2019MNRAS.483.4922B}. However, since its luminosity function at 150\,MHz is mainly obtained by extrapolating observations at higher frequencies, such as 1.4\,GHz and 4.8\,GHz, discrepancies between derived low-frequency simulations and actual low-frequency observations are induced.  

In recent years, the deeper and more extensive low-frequency radio sky surveys \citep{2017MNRAS.464.1146H, 2017A&A...598A.104S, 2019PASA...36...47H,2019A&A...622A...1S,2020PASA...37...18W, 2021PASA...38...14F,2021A&A...648A...2S, 2021A&A...648A...3K, 2021A&A...648A...4D, 2023MNRAS.523.1729B} have provided a wealth of data, significantly enhancing our capacity to refine simulations of the low-frequency radio sky. Among these surveys, the Low-Frequency Array Two-metre Sky Survey (LoTSS) Deep Fields is the most comprehensive deep radio survey ever conducted. In the recent survey by  \citet{2023MNRAS.523.1729B}, the authors provided approximately 80,000 identified radio sources with matched information like stellar mass, star formation rate, etc.. The sources are therefore classified into four categories: HERGs, LERGs, RQ-AGNs, and SFGs. 
Compared to T-RECS, on the one hand, this catalogue offers more diverse sources such as RQ-AGNs, which are not modeled in T-RECS. On the other hand, the catalogue includes sources with higher redshift, which directly removes the discrepancies in source counts at high redshifts \citep{2023MNRAS.523.1729B}.

The forthcoming observations by the Square Kilometre Array (SKA) represent a significant milestone in radio astronomy. The insights obtained through SKA's observations will challenge existing models. Currently, developing more precise simulations is crucial for forecasting SKA observations, addressing the challenges of detecting the EoR signal, and enhancing our understanding of extragalactic point sources.

In this paper, we present an updated classification of radio sources along with their corresponding evolutionary models. The paper is structured as follows: Section~2 outlines the data selection. Sections~3 introduces our models for AGN and SFG. In Section~4, we compare our model outputs with observed results and other simulations. Throughout our analysis, we employ cosmological parameters set to $\Omega_m = 0.3$, $\Omega_{\Lambda} = 0.7$ and $H_0 = 70$ km s$^{-1}$ Mpc$^{-1}$.

\section{ DATA}
As the radio emission is independent of dust, the deep radio survey offers a unique insight into the galaxies and AGN population.
Based on the deep 150\,MHz LoTSS survey in the three distinct sky region: ELAIS-N1, Boötes, and the Lockman Hole,  which have been invested vast amounts of telescope time across the electromagnetic spectrum, 
\citet{2023MNRAS.523.1729B} obtained a catalogue, which contains a total of 81,951 sources, with the lower flux limit of $0.003~\rm mJy$ and the redshift range from $0$ to $7$.
Here we adopted this catalogue to better model the galaxy evolution across cosmic time.

Based on detailed spectral energy distribution fitting, the sources were classified into four categories.  Sources of low radio flux density are classified as SFGs or RQ-AGN depending on whether the spectra are AGN-like or not; while sources of high radio flux density are classified as LERGs and HERGs according to the luminosity. 

Following \citet{2023MNRAS.523.1729B}, the classification of sources is performed by comparing four different spectral energy distribution (SED) codes: MAGPHYS \citep{2008MNRAS.388.1595D}, BAGPIPES \citep{2018MNRAS.480.4379C,2019ApJ...873...44C}, CIGALE \citep{2005MNRAS.360.1413B,2009A&A...507.1793N,2019A&A...622A.103B}, and \textbf{AGNFITTER} \citep{2016ApJ...833...98C}, with the latter two including models that account for contributions from AGN. By evaluating the results from these four codes, they determine whether the sources are radiative mode AGNs and assign an appropriate consensus star formation rate and stellar mass to each source.  Finally, based on whether they are radiative mode AGNs, and using Equation~(2) from \citet{2023MNRAS.523.1729B}—which concerns the relationship between the expected radio emission from the star formation rate and the actual radio emission—the sources are further subdivided into high-excitation galaxies, low-excitation galaxies, star-forming galaxies, and radio-quiet AGNs.

As the extragalactic radio sources primarily consist of AGN and SFGs, this catalogue is ideal for the simulation of foreground sources of the 21\,cm signal. 
Taking into account the incompleteness of faint sources and integrating results from completeness simulations by \citet{2022MNRAS.513.3742K} and \citet{2023MNRAS.523.6082C}, we refined our selection criteria for sources. We focused on those with redshifts between $0$ and $6$ and a flux density threshold of $0.11~\mathrm{mJy}$, which is the lowest flux density at which completeness can be reliably evaluated. This adjusted selection process identified $77,495$ sources, resulting in the exclusion of approximately $3,000$ sources.
%This notable reduction strongly suggests that a significant portion of the excluded sources were those that could not be classified.

\section{MODEL DESCRIPTION}

%Radio sky simulation plays important role in the study of cosmology. Firstly, it can be used to predict the population of radio sources that will be observed in given or future surveys. Secondly, it is useful to assess the completeness of radio survey. Thirdly, it can be used to generate random samples for clustering analyses. Last but most important for us, in the research of reionization, it can be used to predict the foreground noise of the 21\,cm signal.

%The two most widely used radio sky simulations in the literature are the SKA Design Study (SKADS) simulated skies \citep{2008MNRAS.388.1335W} and  the Tiered Radio Extragalactic Continuum Simulation \citep[T-RECS;][]{2019MNRAS.482....2B} 
\subsection{luminosity function}
As the starting point for the radio sky simulation, the measured luminosity functions of different source populations, as well as their cosmic evolution, are the base of the simulations. 
In order to take into account the above with a more complete catalogue given by \citet{2023MNRAS.523.1729B}, we fit the luminosity functions as follows. 
For details not mentions in this chapter (e.g., dark matter simulation for galaxy clustering effect and position angel), we followed what is used in T-RECS model \citep{2023MNRAS.524..993B}.

%AGNs, especially the subset termed radio-loud AGNs, feature radio emission closely related to the accretion of matter by supermassive black holes at the centers of host galaxies. This accretion phenomenon gives rise to the production of narrow, directed jets that extend perpendicular from the plane of the accretion disk \citep{chapman2019foregrounds}.

%In SFGs, the radio emission resembles those of the Milky Way, arising from  synchrotron radiation emitted by relativistic electrons associated with supernovae, as well as free-free emission from HII regions. The observed radio emission from these galaxies often exhibits a strong correlation with their infrared luminosity, which is used to measure the star formation rate (SFR), as referenced in the literature \citep{1992AIPC..254..629C,1985ApJ...298L...7H,2010MNRAS.409...92J}, leading to their classification as Star-Forming Galaxies \citep{chapman2019foregrounds}.

\subsubsection{Luminosity functions for LERGs and HERGs}

We adopted a redshift-dependent comoving luminosity function of double power law as described by \citet{2010MNRAS.404..532M} to represent HERGS and LERGS in units of ${\rm Mpc^{-3}\, (d\log L)^{-1}}$. 
The function is written as

\begin{equation}\label{eq:phi}
\Phi(L, z)=\frac{N}{(L/L_\star)^a+(L/L_\star)^b},
\end{equation}
where $a$, $b$, $L_\star$ and $N$ are the two power indices, the characteristic luminosity and number of sources, respectively. The characteristic quantities of $L_\star$ and $N$ evolve with redshift in the following way:

\begin{equation}\label{eq:N0}
\log N(z) = n_0 + n_1\, \chi +n_{2}\, \chi^2,
\end{equation}

\begin{equation}\label{eq:evol}
L_{\star}(z)\!=\! L_{\star}(0)\,{\rm dex}\!{\left[k_{\rm evo}z\!\left(\!2z_{\rm top}\!-\!{z^{m_{\rm ev}}z_{\rm top}^{(1-m_{\rm ev})\!/(1\!+\! m_{\rm ev})}}\right)\right]},
\end{equation}

\begin{equation}
\chi \equiv \log(1+z),
\end{equation}

\begin{equation}\label{eq:ztop}
z_{\rm top} = z_{{\rm top},0} + \frac{\delta z_{\rm top}}{1 + {L_{\star}(0)}/{L}},
\end{equation}
where $n_0 \equiv \log N(0)$, $N(0)$ and $L_*(0)$ are the characteristic quantities at present or $z=0$, $|m_{\rm ev}| < 1$ is a free parameter, $z_{{\rm top},0}$ and $\delta z_{\rm top}$ are two parameters that very roughly describes the redshift when the luminosity functions reach the top.

\begin{figure*}
	\includegraphics[width=0.9\textwidth]{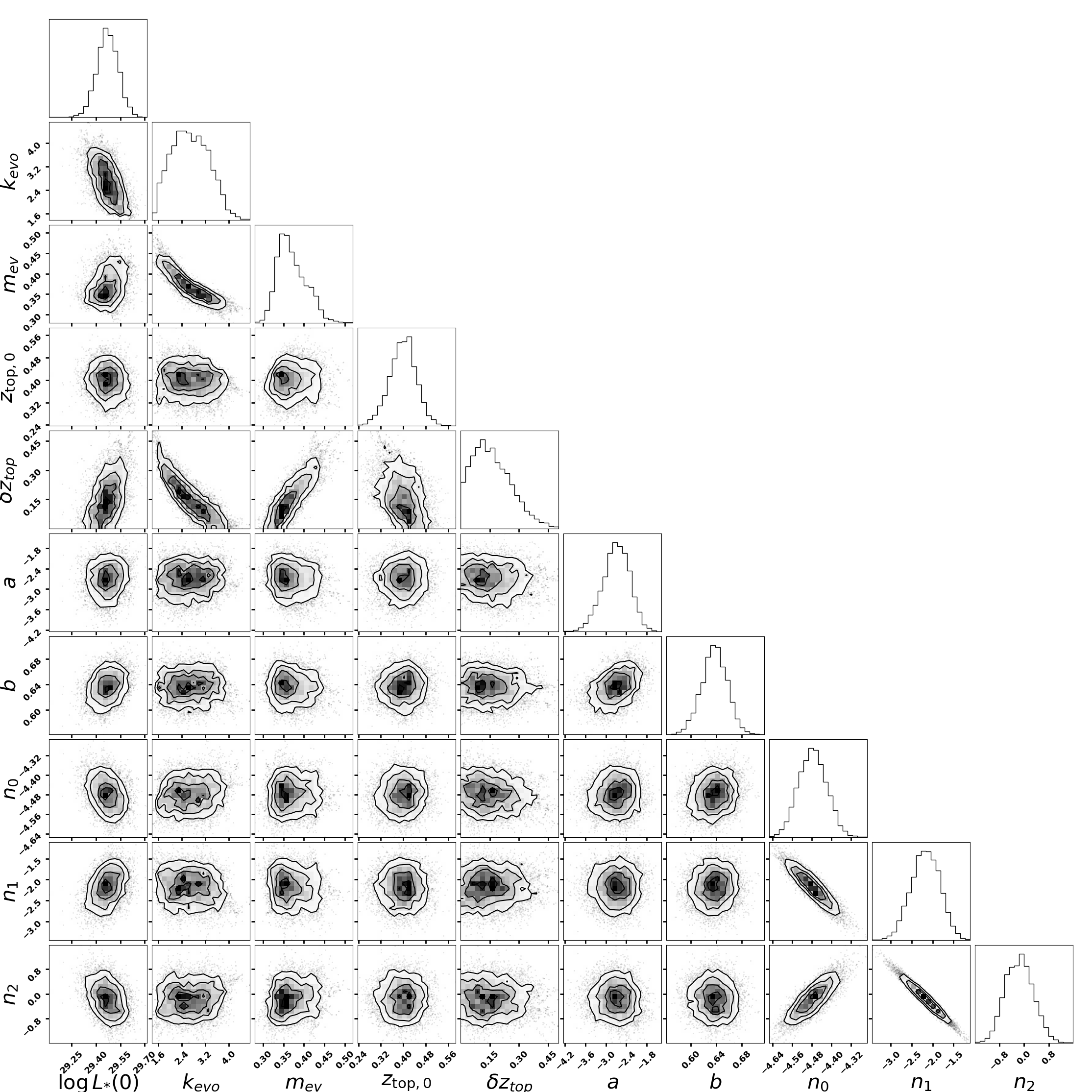}
    \caption{Posterior distributions of the luminosity-redshift evolutionary model for High Excitation Radio Galaxy (HERG) sources.}
    \label{fig:herg_corner}
\end{figure*}

\begin{figure*}
	\includegraphics[width=0.9\textwidth]{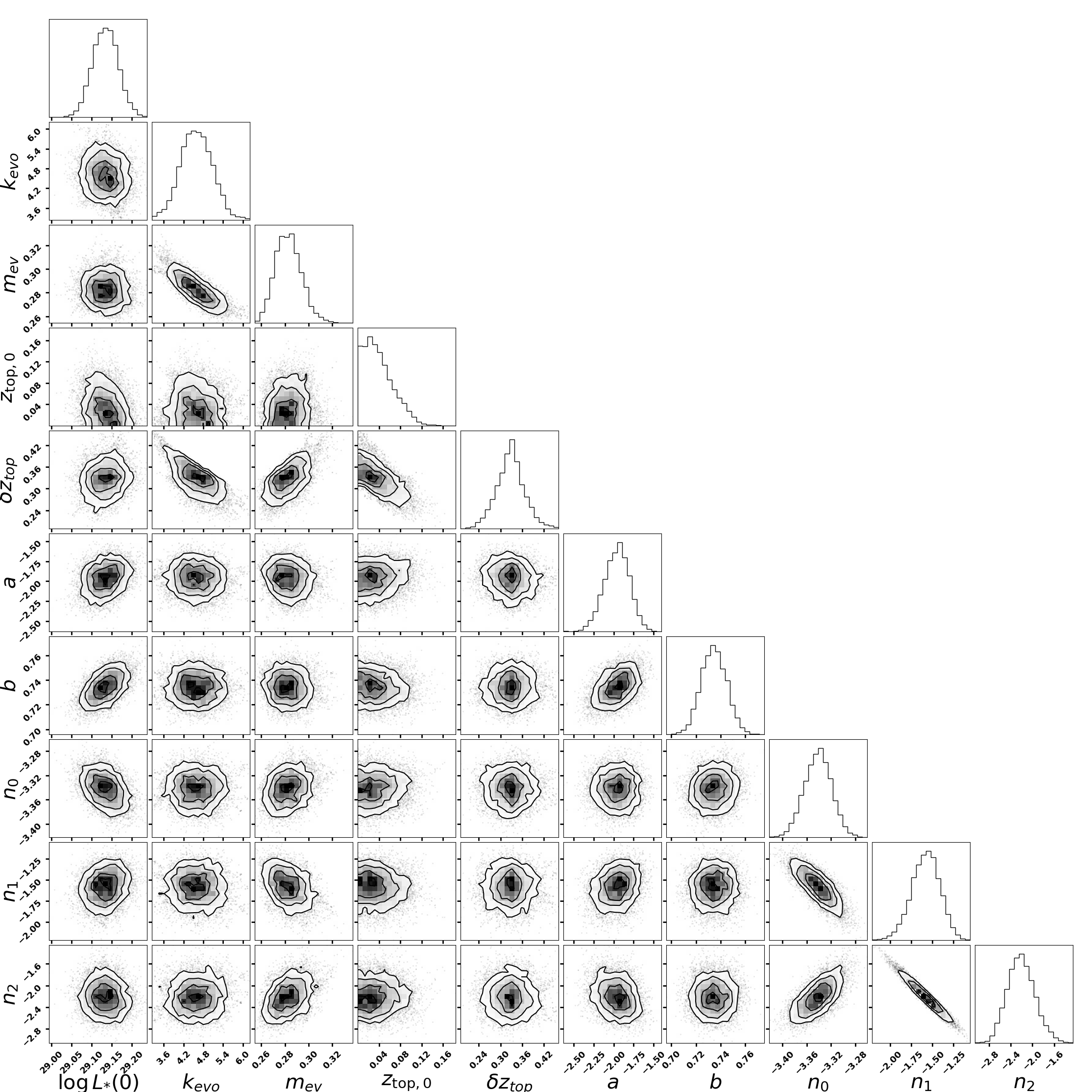}
    \caption{Posterior distributions of the luminosity-redshift evolutionary model for Low Excitation Radio Galaxy (LERG) sources. }
    \label{fig:lerg_corner}
\end{figure*}

Then by fitting the data obtained from the catalogue by \citet{2023MNRAS.523.1729B}, all the parameters in the modelling luminosity functions can be determined.  
Here we use the Monte Carlo Markov Chain method to fit the data, which gives the distribution of probabilities for the parameters. We use the {\it emcee} code to do sampling, which is available online at {\it http://dan.iel.fm/emcee} \citep{2013PASP..125..306F}.
Additionally, during the calculation of the luminosity function for the LoTSS-Deep survey, we take into account the completeness simulations from \citet{2022MNRAS.513.3742K} and \citet{2023MNRAS.523.6082C} for the LoTSS-Deep survey. 
The best-fitting results of the parameters are shown in Table~\ref{tab:agnparameter}, and the contours of the samplings are shown in  Figure~\ref{fig:herg_corner} and Figure~\ref{fig:lerg_corner}.

\begin{table}
 \caption{Best-fit parameters of the evolutionary model for HERGs and LERGs. The luminosity $L_{*}$
 is in $\mathrm{erg\,s^{-1}\,Hz^{-1}}$}
  \begin{tabular}{lccc}
    \hline
    Parameter    & LERG  & HERG \\[5pt]
    \hline
$a$ &  $-1.875^{+0.148}_{-0.157}$   &  $-2.694^{+0.348}_{-0.366}$   \\[5pt]
$b$ &  $0.736^{+0.011}_{-0.010}$   &  $0.636^{+0.021}_{-0.019}$  \\[5pt]
$n_{0}$ &  $-3.341^{+0.022}_{-0.021}$  & $-4.479^{+0.059}_{-0.058}$  \\[5pt]
$n_{1}$ &  $-1.565^{+0.166}_{-0.167}$  & $-2.178^{+0.324}_{-0.341}$  \\[5pt]
$n_{2}$ &  $-2.239^{+0.264}_{-0.245}$  & $-0.151^{+0.433}_{-0.423}$  \\[5pt]
$\log L_{*}(0)$ & $29.137^{+0.029}_{-0.031}$   &  $29.469^{+0.064}_{-0.062}$  \\[5pt]
$k_{\rm evo}$ & $4.459^{+0.528}_{-0.472}$   & $2.616^{+0.683}_{-0.653}$ \\[5pt]
$z_{\rm top,0}$ &  $0.037^{+0.037}_{-0.023}$   &  $0.393^{+0.046}_{-0.052}$ \\[5pt]
$\delta z_{\rm top}$ &  $0.335^{+0.037}_{-0.039}$   & $0.164^{+0.110}_{-0.088}$ \\[5pt]
$m_{\rm ev}$ & $0.284^{+0.011}_{-0.010}$    &  $0.370^{+0.041}_{-0.030}$  \\[5pt]
    \hline
  \end{tabular}  \label{tab:agnparameter}
\end{table}

\begin{figure*}
	\includegraphics[width=0.9\textwidth]{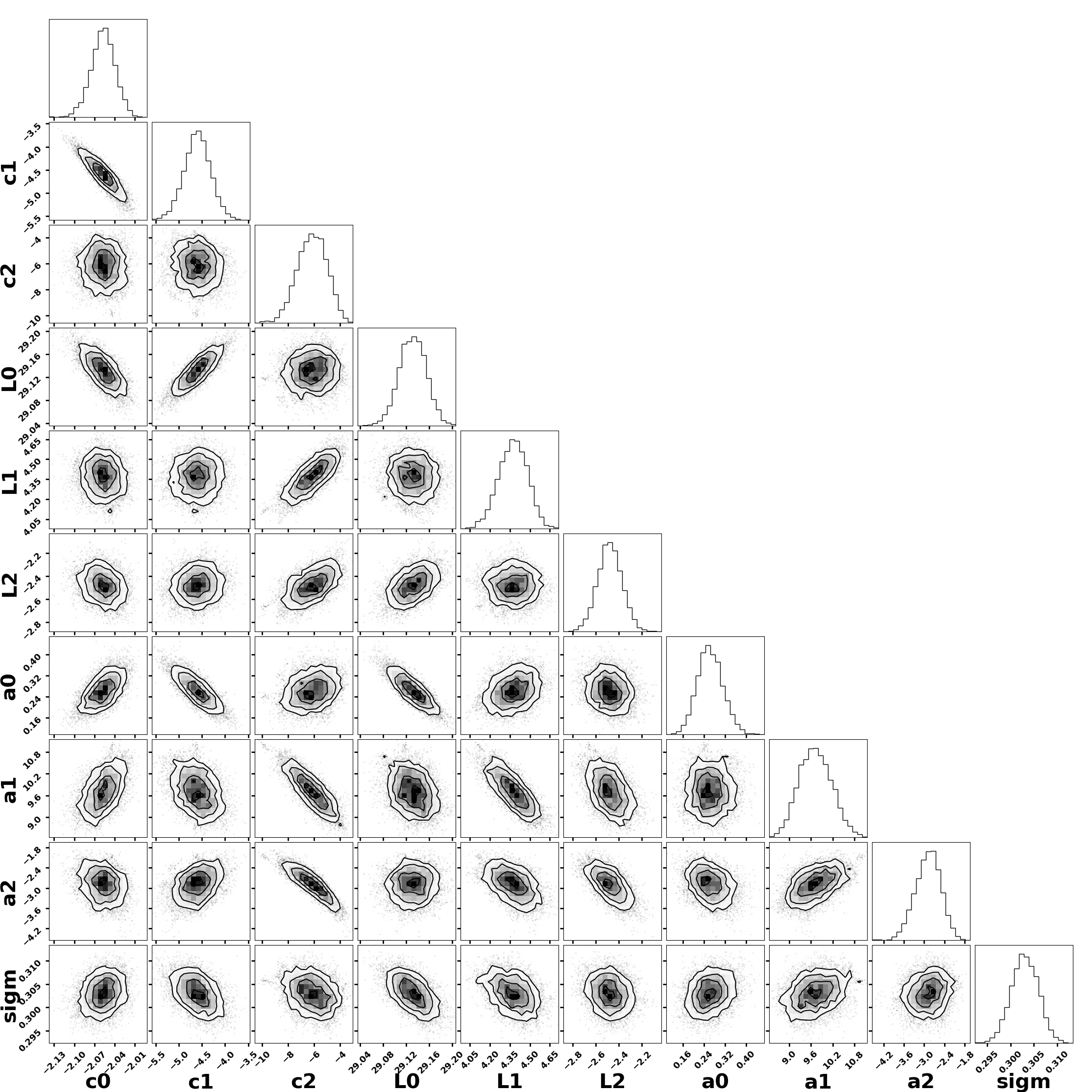}
    \caption{Posterior distributions of the luminosity-redshift evolutionary model for star-forming galaxies (SFGs) sources. }
    \label{fig:sfg_corner}
\end{figure*}

\begin{figure*}
	\includegraphics[width=0.9\textwidth]{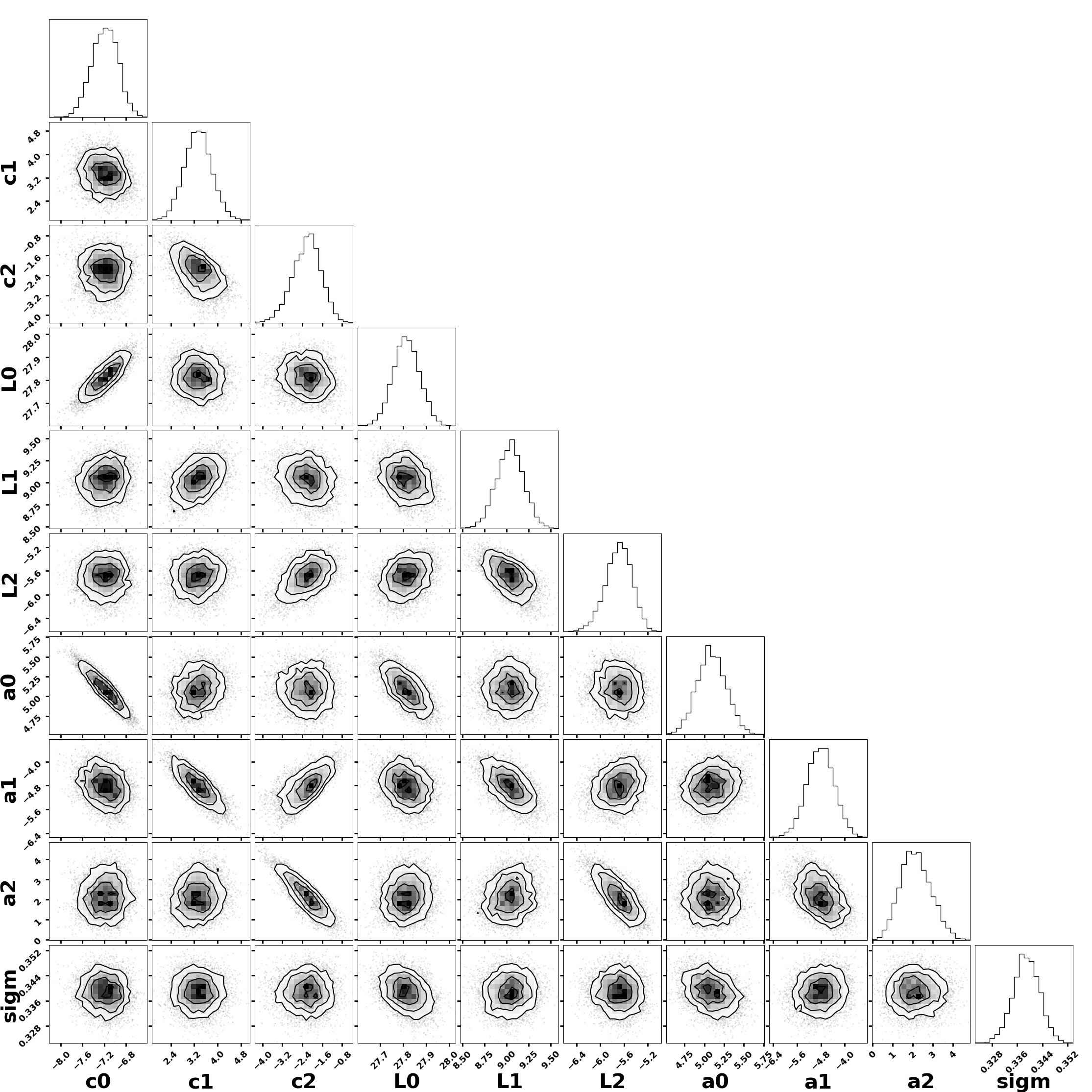
}
    \caption{Posterior distributions of the luminosity-redshift evolutionary model for radio-quiet AGN (RQAGN) sources.}
    \label{fig:rqagn_corner}
\end{figure*}

\begin{figure*}
	\includegraphics[width=0.9\textwidth]{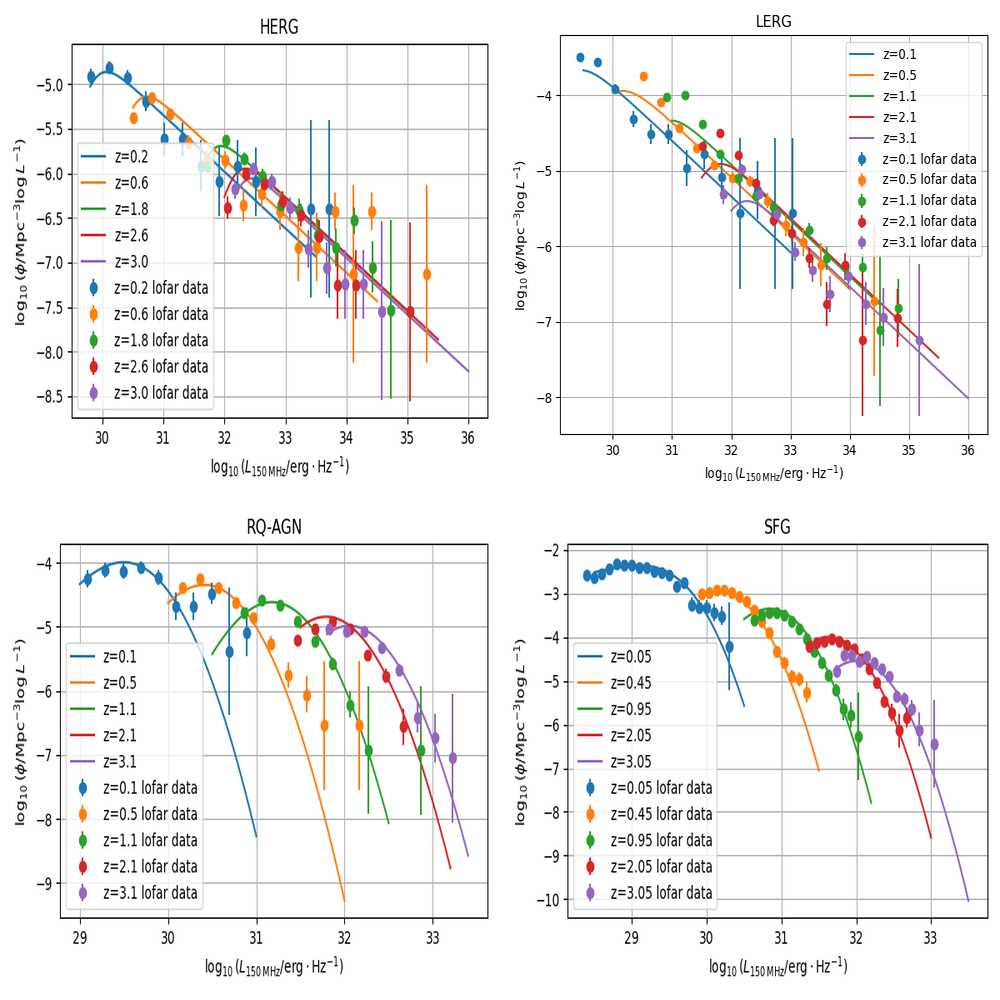}
    \caption{The luminosity functions (LFs) for the HERG, LERG, SFG, and RQ-AGN populations at various redshifts.}
    \label{fig:lf_compare}
\end{figure*}

\subsubsection{Luminosity functions for RQ-AGNs and SFGs}
RQ-AGNs exhibit clear signs of AGN activity across various non-radio wavelengths, including X-rays, mid-infrared, and optical spectra. However, they lack the prominent large-scale radio jets seen in Radio-Loud AGN (RL-AGN), resulting in significantly weaker radio emission \citep{2016ApJ...831..168K, 2016A&ARv..24...13P}. Despite extensive study, the nature of RQ-AGNs remains a subject of ongoing debate. One unresolved question of significance concerns the source of radio emissions in RQ-AGNs. Observational evidence suggests two potential contributors: a central core and star formation processes. Research by \citet{2015MNRAS.452.1263P, 2015MNRAS.453.1079B} suggests that RQ-AGNs exhibit characteristics such as infrared-radio flux ratios, evolving radio luminosity functions, host galaxy colors, optical morphology, and stellar mass similar to those of star-forming systems. This suggests that radio emissions in RQ-AGNs are primarily driven by star formation \citep{2011AJ....141..182K, 2013ApJ...768...37C, 2016ApJ...831..168K}). Conversely, \citet{2015MNRAS.448.2665W, 2017MNRAS.468..217W} contended that RQ-AGNs exhibit higher radio luminosities compared to SFGs with similar stellar masses. \citet{2017ApJ...842...95M} suggested that the conflicting findings may arise partly from variations in luminosity and redshift ranges considered in the studies mentioned above.

%In summary, we adopt the classification method outlined by \citet{2023MNRAS.523.1729B} and 
Considering that the radio emissions of SFGs and RQ-AGNs may have common origins, and recent studies suggest that all types of quasars exhibit a combination of jet and star formation activity \citep{2021MNRAS.506.5888M}, we applied the same function to both RQ-AGNs and SFGs, with the differences attributing to the value of parameters. The luminosity function is taken from a model initially applied to fit the luminosity function of Infrared Astronomical Satellite (IRAS) galaxies \citep{1990MNRAS.242..318S}, which is written as 
\begin{equation}\label{eq:SFRfun}
\Phi(L,z) = C \left( \frac{L}{L_{\star}} \right)^{\alpha} \mathrm{exp} \left\{ -\frac{1}{2} \left[ \frac{\log (1+L/L_{\star})}{\sigma} \right]^{2} \right\} 
\end{equation}

\begin{equation}
C(z)\,  =  C_0 + C_1\, \chi 
+ C_2\, \chi^2  
\end{equation}
\begin{equation}
\log L_{\star}(z)= l_{\star_0} + l_{\star_1}\, \chi 
+l_{\star_2}\, \chi^2 
\end{equation}
\begin{equation}
\alpha(z) = \alpha_0 + \alpha_1\, \chi +\alpha_2\, \chi^2 
\end{equation}

%Where, apart from redshift z and luminosity L, all parameters $C_0$, $C_1$, $C_2$, $L_{\star_0}$, $L_{\star_1}$, $L_{\star_2}$, $\alpha_0$, $\alpha_1$, $\alpha_2$, and $\sigma$ are considered free parameters. 
%We integrate the model with the data and employed {\tt emecc} to fit the free parameters, resulting in updated best-fit results for both RQ-AGN and SFG. Refer to Table~\ref{tab:sfrparameter}, Figure~\ref{fig:sfg_corner}, and Figure~\ref{fig:sfg_corner} for the outcomes.
Similar to the fitting of LERGs and HERGs, the parameters for SFGs and RQ-AGNs are derived by fitting the data from LoTSS with MCMC method. The value of the parameters are shown in Table~\ref{tab:sfrparameter}, and the contours of samplings are shown in Figure~\ref{fig:sfg_corner} and Figure~\ref{fig:rqagn_corner}. In Figure \ref{fig:lf_compare}, we present the fitted luminosity functions at various redshifts for all four types of galaxies as an illustration. \textbf{We note that there is a turnover in the fitted luminosity function at lower luminosities. This turnover has not been reported in previous studies, such as those by \citet{2023MNRAS.523.5292K} for AGN LFs, as these studies excluded low-luminosity data from the LoFAR survey due to concerns about completeness. We have retained the low-luminosity part of LoFAR data in our analysis, which accounts for the observed turnover, particularly for the LERG population. A similar effect can be found for the SFG population when comparing our results to those of \citet{2023MNRAS.523.6082C}. While this turnover influences our predicted source counts at flux densities below the LoTSS observational limits, we emphasize that caution is necessary when extrapolating to significantly lower flux densities, as these predictions may differ substantially from models based on previously published LFs.}

\begin{table}
 \caption{Best-fit results of the parameters of the evolutionary model for RQ-AGNs and SFGs.}
  \begin{tabular}{lccc}
    \hline
    Parameter    & RQ-AGN  & SFG \\[5pt]
    \hline
\textbf{$C_0$} &  \textbf{$-7.190^{+0.236}_{-0.250}$}   &  \textbf{$-2.057^{+0.017}_{-0.018}$} \\[5pt]
\textbf{$C_1$} &  \textbf{$3.327^{+0.462}_{-0.459}$}   &  \textbf{$-4.602^{+0.279}_{-0.286}$}  \\[5pt]
\textbf{$C_2$} &  \textbf{$-2.204^{+0.539}_{-0.614}$}  & \textbf{$-6.142^{+1.107}_{-1.176}$}  \\[5pt]
\textbf{$l_{\star_0}$} & \textbf{$27.812^{+0.058}_{-0.056}$}   &  \textbf{$29.131^{+0.023}_{-0.024}$}  \\[5pt]
\textbf{$l_{\star_1}$} & \textbf{$9.039^{+0.149}_{-0.164}$}   & \textbf{$4.375^{+0.104}_{-0.112}$} \\[5pt]
\textbf{$l_{\star_2}$} &  \textbf{$-5.686^{+0.207}_{-0.233}$}   &  \textbf{$-2.480^{+0.107}_{-0.104}$}\\[5pt]
\textbf{$\alpha_0$} &  \textbf{$5.087^{+0.189}_{-0.182}$}   &  \textbf{$0.262^{+0.049}_{-0.045}$} \\[5pt]
\textbf{$\alpha_1$} &  \textbf{$-4.813^{+0.448}_{-0.450}$}   &  \textbf{$9.711^{+0.471}_{-0.436}$}  \\[5pt]
\textbf{$\alpha_2$} &  \textbf{$2.117^{+0.827}_{-0.721}$}
  & \textbf{$-2.871^{+0.350}_{-0.404}$}\\[5pt]
\textbf{$\sigma$} & \textbf{$0.339^{+0.004}_{-0.004}$}
   & \textbf{$0.303^{+0.003}_{-0.003}$} \\[5pt]
    \hline
  \end{tabular}  \label{tab:sfrparameter}
\end{table}

\subsection{Total intensity number counts}

In the redshift-dependent evolutionary luminosity function model, we take into account the values of $\Phi(L|z)$ for small luminosity intervals $\Delta \log(L)$ and redshift intervals $ \Delta z_i$.  This enables us to calculate the number of sources of each type within a given interval. The formula for this calculation is as follows,
\begin{equation}\label{eq:counts_part}
\Delta N(L)_{i,j} = \Omega \, \Phi(L_j|z_i) \, V( \Delta z) \, \Delta \log(L),
\end{equation}
where $\Omega$ represents the simulated solid angle, $z_i$ denotes the central point of the redshift bin, and $V(\Delta z)$ corresponds to the volume associated with both the solid angle and redshift bin. We assumed
that sources are randomly distributed within $\Delta \log(L)$ and $\Delta z$, and we found when $\Delta z$ = 0.005 and $\Delta \log(L)$ = 0.005 for AGNs and SFGs, the fitting results are the best. The total source count is given by:
\begin{equation}\label{eq:counts_tot}
N = \sum_{i,j} \Delta N(L)_{i,j}.
\end{equation}

\subsection{Source sizes}

%For AGN, we adopt the two models proposed by \citet{2013ApJ...774...24D}. This paper employs Monte Carlo methods to simulate sources with random orientations, generating them from a reasonable intrinsic size distribution. Three sample properties are used as constraints, drawn from \citet{1989ApJ...336..606B}, \citet{2013ApJ...766...37S}, and \citet{2011AAS...21714222B}. By utilizing the observed angular sizes and redshifts of the samples, and setting the distribution of the sources' observing angles to $N(\theta) = \sin\theta$, the simulation is repeated 1000 times. The resulting model incorporates a linear fit for larger intrinsic sizes and a power-law fit for smaller intrinsic sizes, with the actual size of the function's breakpoint determined through fitting \citep{2013ApJ...774...24D}. The generated model is presented in the table 2 of \citet{2013ApJ...774...24D}. Using 40 kpc as the threshold to differentiate between large and small sources, the wide model results in $80\%$ to $90\%$ large sources. In contrast, the narrow model produces a more conservative size distribution estimate, using the same functional form as the wide model, with the proportion of large to small sources being $40\%$-$50\%$ \citep{1984ARA&A..22..319B}. We offer two models to choose from, with the default for AGNs being the more conservative narrow model \citep{2013ApJ...774...24D}.

To model the sizes of HERG and LERG, we adopt the narrow model outlined by \citet{2013ApJ...774...24D} (see their table 2). This approach utilizes Monte Carlo simulations to generate sources with random angular sizes and orientations. The probability distributions for these simulations are informed by the findings from \citet{1989ApJ...336..606B}, \citet{2013ApJ...766...37S}, and \citet{2011AAS...21714222B}. The orientation angle is modeled as $N(\theta) = \sin\theta$, where $\theta$ represents the orientation angle. The model subsequently integrates a linear fit to accommodate larger intrinsic sizes and a power-law fit for smaller sizes, with the breakpoint of the function precisely determined through a fitting process.   
%\textcolor{orange}{Using a threshold of 40 kpc to distinguish between large and small sources, the wide model indicates that large sources make up between $80\%$ to $90\%$ of the population. In contrast, the narrow model, which shares the same functional form as the wide model, tends to provide a more conservative estimate of the size distribution, with the ratio of large to small sources ranging from $40\%$ to $50\%$ \citep{1984ARA&A..22..319B}. We default to the more conservative narrow model for HERGs and LERGs.   Additionally, we adopt the same angular distribution for generating sources as described in this paper, which is $N(\theta) = \sin\theta$.}
%The generated model is presented in the table 2 of \citet{2013ApJ...774...24D}. Using 40 kpc as the threshold to differentiate between large and small sources, the wide model results in $80\%$ to $90\%$ large sources. In contrast, the narrow model produces a more conservative size distribution estimate, using the same functional form as the wide model, with the proportion of large to small sources being $40\%$-$50\%$ \citep{1984ARA&A..22..319B}. We offer two models to choose from, with the default for AGNs being the more conservative narrow model \citep{2013ApJ...774...24D}.

We note that, due to the different classification for SFGs, the latest version of T-RECS \citep{2023MNRAS.524..993B} can not be applied. We therefore adopted the same shape model for SFGs as the old version of T-RECS \citep{2019MNRAS.482....2B}.
%we adopted the same model as the previous version of TRECS\citep{2019MNRAS.482....2B}, although the latest version of TRECS \citep{2023MNRAS.524..993B} has introduced updates to this model.
To be specific, this model determines the stellar mass based on the relationship between dark matter halos and stellar mass ($M_* - M_H$), as presented in table~2 of \citet{2015ApJ...810...74A}. Subsequently, it derives the intrinsic size of the source from the relationship between stellar mass and source intrinsic size according to \citet{2003MNRAS.343..978S}. The related free parameters are determined using results re-fitted by the T-RECS team. Regarding the shape, we employ the ellipticity distribution model from \citet{2016MNRAS.463.3339T}. Ellipticities are randomly assigned to each source according to the distribution model, and the axis ratio is then calculated.

\subsection{Star formation rate}\label{subsec:SFR}

%We have updated the star formation rate (SFR) for \sout{SFGs} \textcolor{orange}{all sources} based on the source classification criteria in \citet{2023MNRAS.523.1729B}.
We updated the star formation rate for all sources using equation 2 in section 6 of \citet{2023MNRAS.523.1729B} and also their source classification method. A statistical relationship between luminosity and star formation rate has been established for sources in different flux intervals, serving as the standard for assigning SFRs. 

\textbf{For RQ-AGN and SFGs, the star formation rate is given by
\begin{equation}\label{eq:sfr_sfg}
{\rm SFR_{SFG}} = \frac{{L - 22.24 }}{{1.08}} + \text{Gaussian}(\mu, v),
\end{equation}
where $L$ represents the radio luminosity of the source, measured in $\mathrm{W\,Hz^{-1}}$. The term \text{Gaussian()} represents a random number generated from a Gaussian distribution, where $\mu$ is the mean and $v$ is the variance both obtained by fitting the sources from LoTSS. }

\textbf{For HERGs and LERGs, it is necessary to account for an additional radio excess, typically around 0.7 dex offset from the standard \( L_{\rm rad} \)-SFR relation described above. The star formation rate for these sources is then given by
\begin{equation}\label{eq:sfr_agn}
{\rm SFR_{AGN}} = \frac{{L - 22.24 - 0.7 - \text{RandomExponential}(\lambda)}}{{1.08}},
\end{equation}
where \text{RandomExponential()} is a random number generator for the exponential distribution, and $\lambda = 0.7952$, which is obtained by fitting the sources from LoTSS. The offset of 0.7 dex arises because both HERGs and LERGs exhibit significant radio excess. Therefore, the overall SFR is lower compared to the relation established for RQ-AGN and SFGs, with an additional random variation accounted for by the exponential distribution. We fit the data from different flux intervals to obtain the parameters shown in Tables \ref{tab:gaussian_rqang} and \ref{tab:gaussian_sfg}.}

\begin{table}
    \centering
    \caption{These are the relevant parameters for the Gaussian fitting in the SFR model for RQAGN.}
    \begin{tabular}{|c|c|c|}
    \hline
    flux range($\mathrm{Jy}$) & mean$\mu$ & variance $v$ \\
    \hline
    $0.0-0.000143$ & $0.0594$ & $0.0655$ \\
    $0.000143-0.000226$ & $-0.0477$ & $0.0582$ \\
    $0.000226-0.000358$ & $-0.179$ & $0.0603$ \\
    $0.000358-0.000568$ & $-0.244$ & $0.0659$ \\
    $0.000568-0.000900$ & $-0.328$ & $0.0640$ \\
    $0.000900-1000.0$ & $-0.353$ & $0.0782$ \\
    \hline
    \end{tabular}\label{tab:gaussian_rqang}
\end{table}

\begin{table}
    \centering
    \caption{These are the relevant parameters for the Gaussian fitting in the SFR model for SFG.}
    \begin{tabular}{|c|c|c|}
    \hline
    flux range($\mathrm{Jy}$) & mean$\mu$ & variance $v$ \\
    \hline
    $0.0-0.000143$ & $0.0302$ & $0.0575$ \\
    $0.000143-0.000226$ & $-0.0463$ & $0.0591$ \\
    $0.000226-0.000358$ & $-0.118$ & $0.0607$ \\
    $0.000358-0.000568$ & $-0.173$ & $0.0624$ \\
    $0.000568-0.000900$ & $-0.198$ & $0.0654$ \\
    $0.000900-1000.0$ & $-0.185$ & $0.0695$ \\
    \hline
    \end{tabular}\label{tab:gaussian_sfg}
\end{table}

\subsection{Spectral index}\label{subsec:alpha}
%For the spectral indices of all sources, we have provided the option to choose the spectral indices. The distribution of spectral indices $\alpha_L$ ($76-227~\rm MHz$) provided by \citet{2021PASA...38...41F} for their sample has been Gaussian fitted, and spectral indices are assigned to sources according to this distribution. 

%\textcolor{orange}{We perform a Gaussian fit on the spectral indices $\alpha_L$(76-227MHz) from the sample provided by \citet{2021PASA...38...41F}, obtaining the distribution of spectral indices.  The spectral indices of all sources are then assigned spectral indices based on this distribution.}

We performed Gaussian fitting on the spectral index distribution, $\alpha_L$, of AGNs using the dataset provided by \citet{2021PASA...38...41F}, resulting in a mean value of -0.701 and a variance of 0.0421. This distribution was then used to generate the spectral indices for HERGs and LERGs.  Following a similar method, we analyzed the spectral index distribution of SFGs from the same dataset, achieving a mean of -0.603 and a variance of 0.0372. We subsequently utilized this distribution to generate spectral indices for RQ-AGNs and SFGs.

%\textcolor{orange}{We performed Gaussian fitting on the spectral index distribution $\alpha_L$ of AGNs from the sample provided by \citet{2021PASA...38...41F}, obtaining a mean of -0.701 and a variance of 0.0421. We then applied this distribution to generate spectral index for HERGs and LERGs. Similarly, we applied the same procedure to the spectral index distribution of SFGs from the sample provided by \citet{2021PASA...38...41F}, obtaining a mean of -0.603 and a variance of 0.0372, and used the distribution to generate spectral index for RQ-AGN and SFG.}

\subsection{Polarization}
%Due to the absence of low-frequency polarization observations, we apply the 1.4\,GHz polarization model proposed by \citet{2014MNRAS.440.3113H} to \textcolor{orange}{LERG and HERG}, similar to the polarization model adopted by T-RECS for steep-spectrum AGN. For the SFG polarization model, we use the model consistent with T-RECS \citep{2012A&A...543A.127S}. Our low-frequency polarization results closely match those of T-RECS, as illustrated in Figure \ref{fig:polarization}, owing to the consistency of the selected model.

\begin{figure}
	\includegraphics[width=\columnwidth]{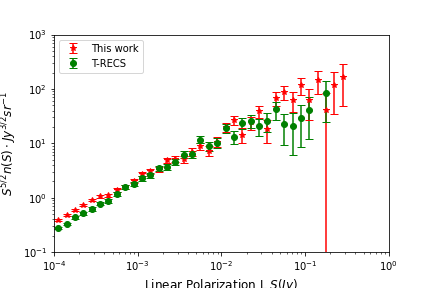}
    \caption{The comparison the differential source counts of polarization in this work with those from T-RECS at 150MHz.}
    \label{fig:polarization}
\end{figure}

Due to the lack of low-frequency polarization observations, we apply the 1.4\,GHz polarization model proposed by \citet{2014MNRAS.440.3113H} to LERGs and HERGs. For the polarization model of SFGs and RQ-AGNs, we use the same model as T-RECS for SFG polarization\citep{2012A&A...543A.127S}. Our low-frequency polarization results closely match those of T-RECS, as illustrated in Figure \ref{fig:polarization}, owing to the consistency of the selected model.

\begin{figure*}
	\includegraphics[width=1.5\columnwidth]{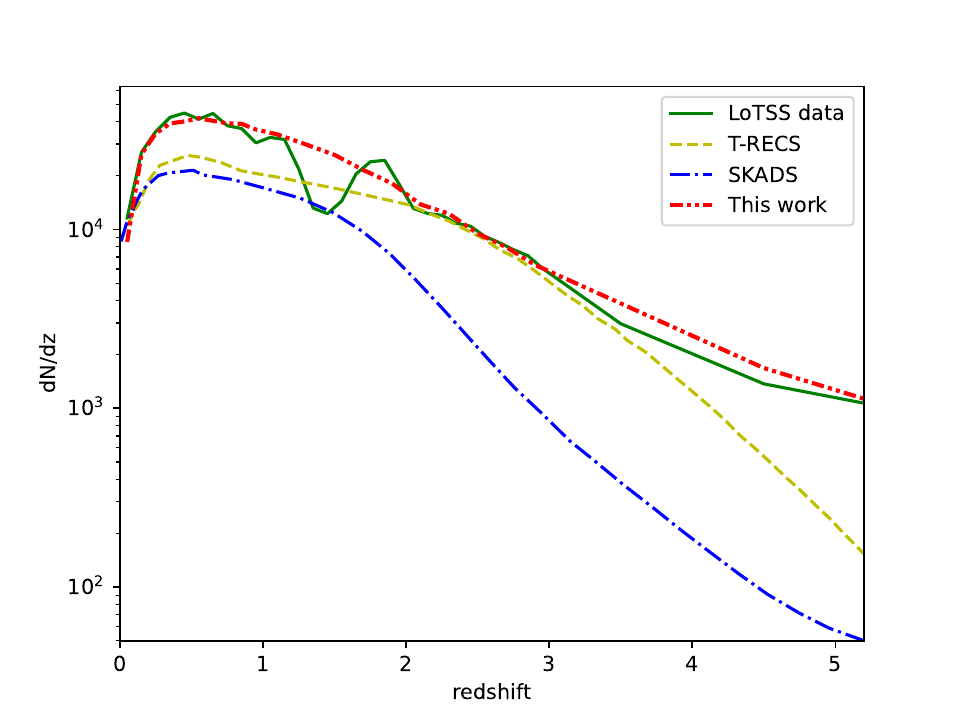}
    \caption{Redshift distribution of radio sources. The green line represents the LoTSS-Deep data. The red  line denotes the result of our model considering the residual sources after completeness simulations from \citet{2022MNRAS.513.3742K,2023MNRAS.523.6082C}. The blue and yellow lines respectively depict the sky simulation predictions of SKADS and T-RECS. }
    \label{fig:dndz}
\end{figure*}

\begin{figure*}
	\includegraphics[width=1.5\columnwidth]{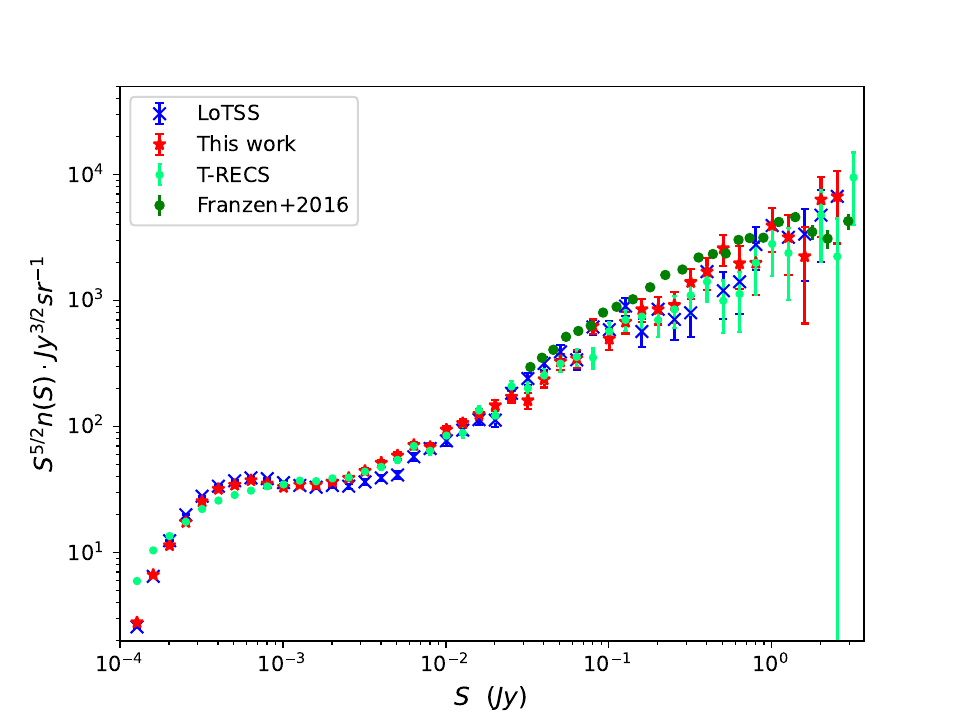}
    \caption{The comparison of differential source counts in total intensity at $150~\rm MHz$. The blue points represent the observational data from LOFAR. The cyan points and red points are the results from T-RECS and this study, respectively, after considering completeness and randomly removing sources. The green points are from \citet{2016MNRAS.459.3314F}.}
    \label{fig:source_counts}
\end{figure*}
 
\section{VALIDATION}
In this section, we validate our results by comparing them with outcomes from prior simulations, notably SKADS and the latest iterations of T-RECS, as well as the observations from LoTSS-Deep and the study conducted by \citet{2016MNRAS.459.3314F} in this section.

For this purpose, we created a simulation with the model described in the previous section, delineated a celestial expanse spanning 25.65 square degrees and established a flux density cutoff at 0.11 mJy. This criterion was selected to align with the parameters of the LoTSS dataset, facilitating straightforward comparison.

For the LoTSS-Deep comparison, we leverage the completeness simulations executed by \citet{2022MNRAS.513.3742K} and \citet{2023MNRAS.523.6082C} for the LoTSS-Deep, to evaluate the detection probability for each simulated source. Following this, we systematically exclude sources from our catalogue as well as the other prior simulations based on these likelihoods, ensuring a rigorous validation when comparing with LoTSS-Deep.

%In this section, we compare our obtained results with real data and \textbf{\uline{the latest version of T-RECS}} 150\,MHz catalogue. \textbf{\uline{We created a sky area covering 25.65 square degrees, setting a flux density threshold of 0.11 mJy.  This choice ensures consistency with the LoTSS data, facilitating straightforward comparison.}} \textcolor{magenta}{Please make sure you use the latest T-RECS version from Git to produce the T-RECS plots. From Bonaldi et al. 2019 to Bonaldi et al. 2023 there was a recalibration specific to improve the 150 MHz consistency (see fig 6 of the old paper vs Figure 4 of the new one, top left panels. } Initially, we use the completeness simulation carried out by \citet{2022MNRAS.513.3742K} and \citet{2023MNRAS.523.6082C} for LoTSS-Deep to compute the likelihood of each simulated source being observed. Subsequently, we randomly remove sources from the catalogue according to these probabilities.

%\textcolor{magenta}{What sky area did you use for the simulations? this affects the size of error bars in your counts and distribution plots, which are due to sample variance. Also you could quote the flux limit you used for the simulations. } 

We first focus on the redshift distribution, as shown in Figure~\ref{fig:dndz}. For this specific comparison, we employ a flux limit of 0.15 mJy with all types of AGNs and SFGs included. The LoTSS-Deep data, represented by the green line, indicates a notable dip in the redshift range of $1.0 < z < 1.5$. This reduction is attributed to an aliasing effect in the photometric redshifts, due to the absence of H-band data, a topic elaborated upon by \citet{2023MNRAS.523.1729B} and further discussed in \citet{2023MNRAS.523.6082C}. Consequently, we omitted this segment of data from the fitting analysis presented in the preceding section. The red line illustrates the outcomes after the exclusion of certain sources, in alignment with the completeness simulation studies conducted by \citet{2022MNRAS.513.3742K} and \citet{2023MNRAS.523.6082C}. The simulated celestial projections from SKADS and T-RECS are depicted by the blue and yellow lines, respectively, with both models incorporating considerations of LoTSS-Deep completeness. Notably, the SKADS predictions manifest a significant shortfall in source counts beyond a redshift of $z>2$, a trend that T-RECS also exhibits beyond $z>4$. In contrast, our results demonstrate a closer approximation to the empirical observations of the celestial sphere.

%The redshift distribution of radio sources is shown in Figure~\ref{fig:dndz}. \textbf{\uline{Here, we specifically selected sources with flux densities exceeding 0.15 mJy.}} The green line represents the LoTSS-Deep data, where the slight decrease in the range of $1.0 < z < 1.5$ is caused by the confusion effect in the luminosity redshift, which may be due to the lack of H-band data \citep{2023MNRAS.523.1729B}; this is discussed in more detail in \citet{2023MNRAS.523.6082C}, so we excluded this part of the data when fitting. The red line is the result after removing some sources considering the completeness simulation of \citet{2022MNRAS.513.3742K} and \citet{2023MNRAS.523.6082C}. The blue and yellow lines are predictions of the simulated sky from SKADS \citep{2008MNRAS.388.1335W} and T-RECS respectively, both of which also consider completeness. It can be seen that when $z>2$, the number of sources predicted by SKADS is significantly less than the observation, and T-RECS also has the same situation when z>4, and our results are closer to the real sky.

The comparative analysis of differential source counts in total intensity at $150~\mathrm{MHz}$, encompassing AGN and SFG, is illustrated in Figure~\ref{fig:source_counts}. The observations from LOFAR are denoted by blue dots, while the results from T-RECS and this study (both post-adjusted for completeness by randomly subtracting sources) are represented by cyan and red dots, respectively. Observations from \citet{2016MNRAS.459.3314F} are indicated by green dots. It can be found that in the higher flux density spectrum, both T-RECS and our study align closely with the observational data. However, at the lower flux density threshold, near $10^{-4}$ Jy, our study demonstrates a more precise alignment with the observational data compared to T-RECS, highlighting the effectiveness of our adjustments.We note that our analysis is focused at 150\,MHz only, while the T-RECS model covers the 150\,MHz--20\,GHz frequency range. While our model represents an improvement with respect to the original T-RECS model at the low-frequency end, consistency with observations at higher frequencies has not been addressed here.

%The comparison of differential source counts in total intensity at $150~\rm MHz$ is depicted in Figure~\ref{fig:source_counts}, encompassing both AGN and SFG. The blue dots represent LOFAR observational data, while the cyan and red dots illustrate T-RECS and the results from this study after randomly subtracting sources considering completeness.  The green dots showcase the observational results from \citet{2016MNRAS.459.3314F}.  
%[The yellow line represents the AGN model from \citet{2017MNRAS.469.1912B}.]  Notably, in the high flux density range, both T-RECS and our results closely match the observational data.  However, in the low flux density range near $10^{-4}$ Jy, our results exhibit a closer proximity to the observational data compared to T-RECS.

%\subsection{Discussion}
%\textcolor{magenta}{Here you need to state that you are computing power spectra of image cubes with mock point sources to assess foreground contamination to EoR.}

Now we focus on the power spectrum of the image cubes of point sources. To mimic the simulation along the frequency axis, we employ a power-law model with randomly distributed spectral indices, whose distributions are sourced from the sample provided by \citet{2021PASA...38...41F}, as we discussed earlier in \ref{subsec:alpha}. This approach enables us to generate a three-dimensional cube, spanning a frequency range from $150~\mathrm{MHz}$ to $160~\mathrm{MHz}$, with a frequency resolution of $0.02~\mathrm{MHz}$. To model the morphology of the point sources, we utilize $galsim$ \citep{2015A&C....10..121R}, generating images that cover a field of view (FoV) of 100 square degrees for each of the 501 frequency slices. These slices are then merged to form a comprehensive data cube.
The spatial distribution of these point sources is designed to reflect clustering effects\citep{2019MNRAS.482....2B,2019MNRAS.483.4922B,2014MNRAS.440.2115J}. This methodological approach ensures a realistic representation of point source clustering across the simulated data cube.
%We employ a model featuring randomly distributed spectral indices to produce $50$ sets of catalogues, incorporating clustering effects. These catalogues cover a frequency range from $150~\rm MHz$ to $160~\rm MHz$, with a frequency interval of $0.02~\rm MHz$. Using $galsim$ \citep{2015A&C....10..121R}, we generate images of FoV for these 501 slices and amalgamate them into a data cube. Subsequently, we compute both one-dimensional and two-dimensional power spectra for these data cubes. 
For comparison, we also calculated the power spectrum for the T-RECS model as well as the SKADS model \citep{2008MNRAS.388.1335W} using the same FoV and frequencies. For our analysis, we excluded all sources exceeding a brightness temperature of 50,000 K to facilitate comparisons, as these extremely bright sources can disproportionately influence the power spectrum.
%Additionally, we conduct identical calculations for T-RECS models to discern discrepancies in power spectra. Furthermore, we calculate the power spectrum for the catalogue provided by \citet{2008MNRAS.388.1335W} to facilitate comparison.

\begin{figure}
	\includegraphics[width=\columnwidth]{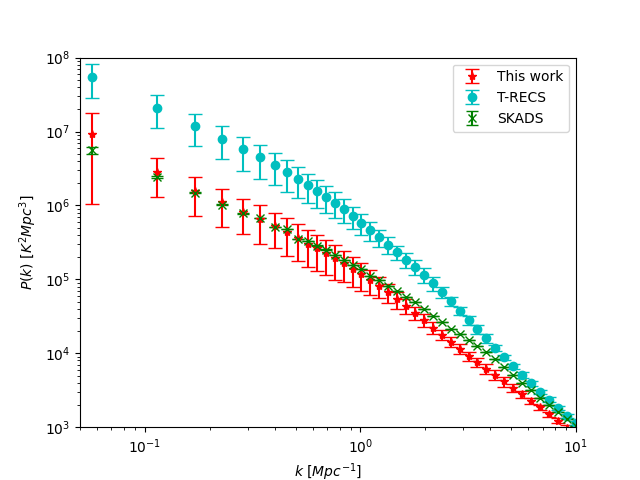}
    \caption{The one-dimensional power spectrum for comparison, all accounting for clustering effects. Red dots denote our model. Cyan dots represent calculations conducted with the T-RECS model. Green dots indicate results obtained using the SKADS model.}
    \label{fig:3d_ps_all}
\end{figure}

For the purpose of error estimation, we conducted fifty simulation sets according to our methodology, and an equivalent number following the T-RECS approach. It's important to note that we do not account for errors in the SKADS simulation due to the availability of only one published dataset.
Figure~\ref{fig:3d_ps_all} presents the averaged results from fifty sets of one-dimensional power spectra, calculated subsequent to the exclusion of notably bright sources. The variance is denoted through error bars. Red points illustrate the calculations derived from our simulation, while the one-dimensional power spectra obtained from the T-RECS and SKADS models are represented by cyan and green points, respectively. Upon comparison, the power spectrum from T-RECS is observed to be higher than that of our model. This discrepancy could be attributed to the inherent differences between the models, where T-RECS incorporates a larger quantity of faint sources compared to our simulation.

\section{Conclusions}

In refining our source classification and luminosity function based on LOFAR observations, we have achieved a more accurate alignment with the redshift distribution and total intensity count of the sources compared to actual data.  
%Furthermore, we have introduced enhancements to the morphology and polarization of the sources, utilizing polarization information from $1.4~\rm GHz$ data due to the limited availability of low-frequency observations. 
%We will update relevant content when related observation results become available. 
We conducted comparisons between our model, actual observations, and other existing models, focusing on critical metrics such as the redshift distribution of radio sources, differential source counts in total intensity, and power spectra. In terms of differential source counts at total intensities below $1~\mathrm{mJy}$, our model shows better alignment with observational data than the T-RECS model at 150 MHz. Moreover, while T-RECS presents higher power spectra—likely due to the inclusion of more faint sources—our model’s power spectra are more closely aligned with those from the SKADS model. Our model also demonstrates improved agreement in the distribution of source counts across redshifts, particularly at higher redshifts, compared to observations from LoTSS. Additionally, our model  allows for the extension of simulations to larger sky areas with lower flux density thresholds, despite the smaller surveyed area of LoTSS. We note that our focus is on low frequencies, and consequently, we did not verify consistency at higher frequencies as performed by the T-RECS model.

We have made the modified T-RECS code, which incorporates our model for simulating low-frequency radio point sources, available to the research community\footnote{https://github.com/nxdtxdyka/f150\_ps\_simulation}. This effort significantly enhances the accuracy of simulations and facilitates further studies in the dynamics of the low-frequency radio source sky.

\section*{Acknowledgements}

We thank ChatGPT for polishing our English in this paper. This work is supported by National SKA Project of China (No. 2020SKA0110100), and the National Natural Science Foundation of
China (No. 12203085, 11925301, 11890692, 11873074, 11525312, 11333005, U1831205, U1531130, Y845281001). HYS acknowledges the support from Key Research Program of Frontier Sciences, CAS, Grant No. ZDBS-LY-7013 and Program of Shanghai Academic/Technology Research Leader. 

%%%%%%%%%%%%%%%%%%%%%%%%%%%%%%%%%%%%%%%%%%%%%%%%%%
\section*{Data Availability}

The data underlying this article will be shared on reasonable request to the corresponding author.
 
%   Additionally, we plan to share the program on GitHub, and the URL for the repository is \url{https://github.com/nxdtxdyka/f150_ps_simulation}.  The format and content of the catalogue generated by our program closely resemble that of T-RECS, as detailed in Table \ref{tab:catalog}.



\bibliographystyle{mnras}
\bibliography{example} % if your bibtex file is called example.bib

%\begin{thebibliography}{99}
%\end{thebibliography}

%%%%%%%%%%%%%%%%%%%%%%%%%%%%%%%%%%%%%%%%%%%%%%%%%%

%%%%%%%%%%%%%%%%% APPENDICES %%%%%%%%%%%%%%%%%%%%%

\appendix

\bsp	% typesetting comment
\label{lastpage}
\end{document}